\documentclass[amssymb,twocolumn,aps]{revtex4}
\usepackage{graphics,color,graphicx,amsmath}
\usepackage{subcaption}
\usepackage{natbib}
\usepackage{verbatim}
\usepackage{graphicx}
\usepackage[above,below]{placeins}
\usepackage{float}
\usepackage{tabularx}
\usepackage{times}
\usepackage{blindtext}
\usepackage{url}
\usepackage{rotating}
\usepackage{mathtools}

\renewenvironment{quote}
  {\list{}{\rightmargin=.1cm \leftmargin=.3cm}%
   \item\relax}
  {\endlist}

\begin{document}

\title{Investigating peer recognition across an introductory physics sequence: Do first impressions last?}

\author{Meagan Sundstrom~\footnote[1]{Corresponding author: mas899@cornell.edu}$^{1,2}$ and Logan Kageorge~\footnote[2]{lkageorge@brenau.edu}$^3$}
\affiliation{$^1$Laboratory of Atomic and Solid State Physics, Cornell University, Ithaca, New York 14853, USA\\
$^2$Department of Physics, Drexel University, Philadelphia, Pennsylvania 19104, USA\\
$^3$Department of Mathematics and Science, Brenau University, Gainesville, Georgia 30501, USA}

\date{\today}

\begin{abstract}

Students' beliefs about the extent to which meaningful others, including their peers, recognize them as a strong science student are correlated with their persistence in science courses and careers. Yet, prior work has found a gender bias in peer recognition, in which student nominations of strong peers disproportionately favor men over women, in some instructional contexts. Researchers have hypothesized that such a gender bias diminishes over time, as determined by students' academic year: studies have found a gender bias in science courses aimed at first-year students, but not in science courses aimed at beyond first-year students. This hypothesis that patterns of peer recognition change over time, however, has yet to be tested with longitudinal data--previous studies only examine snapshots of different students in different science courses. In this study, we isolate the effect of time on peer recognition by analyzing student nominations of strong peers across a two-semester introductory physics course sequence, containing the same set of students and the same instructor in both semesters, at a mostly-women institution. Using a combination of social network analysis and qualitative methods, we find that while many students receive similar levels of peer recognition over time, the four most highly nominated students--the recognition celebrities--exhibit some change between semesters even in this highly controlled setting. Furthermore, we observe that these changes in the celebrities track closely with changes in student outspokenness and that being outspoken is likely more important for gaining recognition than earning a high grade in the class. 
These findings lend support to prior work's hypothesis that peer recognition changes over time, but also challenge the generalizability of previous results (i.e., that patterns of recognition are related to students' academic year). Instead, peer recognition seems highly sensitive to variables such as individual students' participation and, therefore, may be course-specific. We provide recommendations for both when and how instructors may intervene on peer recognition based on our results. 




\end{abstract}

\maketitle

\section{Introduction}

An individual's \textit{identity} refers to their being a ``certain kind of person" in a given context~\cite{gee2000chapter}. Thus, one's \textit{science identity} is the extent to which an individual believes they are a ``science person." 
Researchers have modeled science identity as containing three dimensions: performance and competence, interest, and recognition~\cite{carlone2007understanding,hazari2010connecting}. Performance and competence refers to student beliefs in their abilities to perform science tasks and understand science content. Interest is students' enjoyment or interest in engaging with science topics. Recognition is synonymous with students' being perceived by meaningful others, including their peers, instructors, friends, and family, as a science person. While all three dimensions positively correlate with student outcomes, such as participation and persistence in their science courses and their intentions to pursue a scientific career, studies have shown that \textit{recognition} is the dimension that most strongly correlates with such outcomes~\cite{hyater2018critical,lock2013physics,carlone2007understanding,hazari2010connecting,hazari2018towards,hazari2017importance,kalender2019gendered,boe2023cleverness}. When students feel recognized by other people as a science person, they are more likely to perceive themselves as a science person and subsequently persist in their science course and intend to pursue a scientific career~\cite{hazari2018towards, kalender2019female}.

\textit{Peer recognition} is of particular importance for undergraduate students because they interact with and observe their classmates often. Research has shown, however, that student recognition of their strong science peers often exhibits a gender bias in which men disproportionately receive more recognition from their peers than women~\cite{grunspan2016,salehi2019,bloodhart2020,sundstrom2022perceptions,sundstromWhoWhat}. Such a gender bias may lead  women to feel less recognized by their peers as strong science students than men, putting women at a disadvantage for developing their science identities and persisting in science. 

In one such study, Grunspan and colleagues investigated large, introductory biology courses containing mostly first-year students~\cite{grunspan2016}. In each course, the authors observed that men received significantly more nominations as strong in the course material than women. Salehi and colleagues~\cite{salehi2019} later performed similar work for two mechanical engineering courses taken by second and third-year students. The authors found no gender bias in student nominations of strong peers in either course. Bloodhart and colleagues~\cite{bloodhart2020} then analyzed peer recognition across many introductory life science and physics courses. Students in the life science courses were mostly first-year students and the physics courses were spread across all four academic years. The researchers found in both disciplines that both men and women undernominated women as knowledgeable in the course material. Additionally, our prior work investigated peer recognition in three remote, introductory physics courses~\cite{sundstrom2022perceptions}. We observed a gender bias in student nominations of strong peers in the two courses aimed at first-year students, but not in the third course aimed at second-year students. Another study of ours found a gender bias in peer recognition in four different introductory physics courses containing a majority of first-year students~\cite{sundstromWhoWhat}.

Comparing across these previous studies~\cite{grunspan2016,salehi2019,bloodhart2020,sundstrom2022perceptions,sundstromWhoWhat}, the presence or absence of gender bias in peer recognition is more related to whether the course contains a majority of first-year or beyond first-year students than other course features, such as scientific discipline and student demographics~\cite{sundstrom2022perceptions}. Specifically, these studies observed that gender bias in peer recognition is consistently present in science courses aimed at first-year students, but not in science courses aimed at beyond first-year students~\cite{grunspan2016,salehi2019,bloodhart2020,sundstrom2022perceptions,sundstromWhoWhat}. Researchers, therefore, have hypothesized that patterns of peer recognition change over the time-scale of semesters and years, becoming less biased due to students' increased familiarity with diverse peers~\cite{sundstrom2022perceptions}.  

These previous studies, however, only examined snapshots of peer recognition among different sets of students in different science courses and so do not provide direct evidence of change in peer recognition over time~\cite{grunspan2016,salehi2019,bloodhart2020,sundstrom2022perceptions,sundstromWhoWhat}. To our knowledge, no research has investigated students' recognition of their strong peers over a time-scale longer than one semester, such as by following the same set of students throughout their course sequence. Yet such an understanding is crucial for identifying when instructional interventions, such as those aimed at equitable peer recognition, are likely to be most effective. If, for example, biases in peer recognition primarily occur early on in a course sequence as in prior work~\cite{grunspan2016,salehi2019,bloodhart2020,sundstrom2022perceptions,sundstromWhoWhat}, instructional interventions ought to be implemented early on to mitigate possible undesirable effects (e.g., women disproportionately not enrolling in subsequent courses due to lack of peer recognition). On the other hand, if biases in peer recognition do not follow a consistent pattern over time, interventions may need to be implemented consistently throughout a course sequence.

In the current study, therefore, we examine patterns of peer recognition for a single cohort of students across their introductory physics course sequence. We also conduct our study in a unique, understudied context in physics education research~\cite{cid2020demographics}: a mostly-women, liberal arts college. In doing so, we control for multiple variables that may shape peer recognition: instructor (the instructor was the same in both semesters of our study), instructional style (the same pedagogy was used in both semesters), fluctuating student enrollment between semesters (all students enrolled in the spring course were also in the fall course), student gender (almost all students included in the analysis identified as women), and student academic year (all students were beyond their first year). Thus, our research design largely isolates the effect of time on patterns of peer recognition, allowing us to robustly test prior work's hypothesis that peer recognition changes between semesters.

In our study, we first aim to determine whether and how the students receiving the most recognition from their peers--the recognition \textit{celebrities}~\cite{grunspan2016}--change between the two semesters of the course. We also aim to understand how the reasons for which students nominate their strong peers change over time by building on two existing threads of research: the relationship between peer recognition and student outspokenness and the relationship between peer recognition and student performance~\cite{grunspan2016,salehi2019,bloodhart2020,sundstrom2022perceptions,sundstromWhoWhat}.


\subsection{Peer recognition and outspokenness}

Previous studies have found a strong correlation between students' outspokenness, the extent to which they verbally participate in class, and their peer recognition, or how many nominations they receive from other students as strong in the course~\cite{grunspan2016,sundstrom2022perceptions}. In particular, this research has demonstrated that a gender disparity in outspokenness coincides with gender bias in peer recognition: when men disproportionately participate or speak up more than women in class, there tends to be a gender bias in peer recognition favoring men over women. Outspokenness has also been measured in different ways. Grunspan and colleagues asked the course instructor which students were the most active participants at the end of each class session~\cite{grunspan2016}, while in our prior study we quantified students' participation in online discussion boards during remote courses~\cite{sundstrom2022perceptions}. 
In the current study, we expand on this body of work by examining the role of outspokenness in peer recognition throughout a two-semester course sequence. 

Our previous work also identified that peer recognition forms in two ways: direct interactions with peers (e.g., collaborating on a homework assignment) and indirect observations of peers (e.g., watching a near-peer correctly solve a physics problem)~\cite{sundstromWhoWhat}. Correspondingly, outspokenness may occur within peer interactions (e.g., during groupwork activities or out-of-class study groups) or through other means (e.g., offering answers to instructor questions during lectures). In this study, we disentangle the effects of these two different kinds of outspokenness on peer recognition over time by directly comparing peer recognition to peer interactions.

\subsection{Peer recognition and academic performance}

Prior work has uniformly found that students who earn higher grades in their science course tend to receive more nominations from their peers as strong in the course~\cite{grunspan2016,salehi2019,sundstrom2022perceptions,sundstromWhoWhat}. The studies that more closely examined the individual students receiving the most peer recognition, however, observed that high course grades are not sufficient for becoming a celebrity~\cite{grunspan2016,salehi2019}. That is, students receiving the most peer recognition are not necessarily the highest performers in the class. This effect was demonstrated in both first-year and beyond first-year science courses, yet it remains unclear if such a pattern is consistent over time for a given set of students. Thus, we investigate whether and how the relationship between peer recognition and academic performance changes over time for a single cohort of physics students.

\subsection{Current study}

To summarize, researchers have found a gender bias in peer recognition in first-year, but not beyond first-year, science courses. This body of work suggests that peer recognition changes over time as students become more familiar with a diverse set of their peers. Researchers, however, have not yet examined peer recognition among a particular set of students on a time-scale longer than one semester. 
Relatedly, the roles of outspokenness and academic performance in peer recognition over time are not well understood. In the current study, therefore, we investigate the effect of time on who and what gets recognized in peer recognition by examining a single cohort of students across a two-semester introductory physics course sequence. We aim to address the following research questions: 
\begin{enumerate}
    \item How does students' received peer recognition change throughout an introductory physics course sequence?
    \item How do the reasons for which students perceive their strong peers change throughout an introductory physics course sequence?
\end{enumerate}

\section{Methods}

In this section, we first describe the instructional context of our study and then provide details about our data collection and analysis methods.

\subsection{Instructional context}

The data for this study were collected at Brenau University, a small, private, primarily undergraduate institution (PUI) in Gainesville, Georgia. The undergraduate population at Brenau is non-traditional compared to similar PUIs. According to the university's Institutional Research Office, which only collects and stores data on student gender as binary, 89\% of undergraduates identify as women and 11\% of undergraduates identify as men. In addition, racial demographics mimic those of the state as a whole, with White students comprising less than 50\% of the student population and Black and Hispanic students accounting for more than 35\% of the student population. Brenau also boasts a high age diversity, with less than 25\% of students within traditional college age (18--21 years old).

This study investigates a single cohort of students through an in-person, two-semester, algebra-based introductory physics course comprised of a mechanics course in the fall and an electricity and magnetism course in the spring. These are the only physics courses offered at the institution, as Brenau does not support a physics major. All students enrolled in the spring electricity and magnetism course were also enrolled in the fall mechanics course (see Table~\ref{tab:demographics}). The total enrollment dropped by three students who did not register for the spring course, either for personal reasons or because the course was not required by their major.

Both courses were taught by the same instructor (the second author of this paper) and were presented as a flipped classroom, with students watching prepared lecture videos and completing homework assignments before attending twice-weekly lectures and weekly laboratory (lab) or group problem  (GP) sessions. Time in lecture was spent reviewing the course material with an emphasis on student problem solving and peer interactions. In lab, students completed an experiment in groups of three to four and then prepared video lab reports outside of class, similar to the lab curriculum described in Ref.~\cite{lin2014peer}.
In GPs, students worked in their lab groups to complete more challenging problem sets with occasional instructor guidance.

The grading scheme was the same in both courses and grades were largely determined by the one course instructor. Final course grades were mostly based on exams (30\%), lab and GP participation (20\%), lab presentations (15\%), lecture participation (10\%), and homework (10\%). Exams emphasized similar problem solving skills that students practiced during GP sessions and were completed individually. Exams were take-home to allow for greater flexibility and accessibility, and were graded by the instructor. Lab and GP participation accounted for students actively conducting lab activities and engaging in solving physics problems in their small groups during class. These participation grades were determined by the instructor. Lab presentations were delivered by lab groups to the whole class after each experiment and were peer-graded. Students were trained to use a rubric to grade other groups' presentations using criteria such as overall structure, description of the model being tested, and description of the performed experiment. Students also assessed their own groupmates on criteria such as contribution, communication, dependability, and helpfulness. Each student received their own lab presentation grade based on a combination of their grades earned from students outside of their group and their grades earned from their groupmates. Lecture participation was graded based on responses to live poll questions during class. Students were given credit for completion, determined by the instructor. Homework assignments consisted of watching short videos and answering related questions before each lecture. Homework was completed individually and could be attempted multiple times. These assignments were auto-graded by the learning management system, where the correct answers were input by the instructor. 


\begin{table}[t]
\caption{\label{tab:demographics}%
Summary of survey respondents. All students in the spring electricity and magnetism course were also in the fall mechanics course. Most demographic variables are considered binary to preserve anonymity. ``Other" race or ethnicity indicates students who identify as at least one of the following: American Indian or Alaska Native, Asian or Asian American, Black or African American, and Hispanic. SD indicates standard deviation.
}
\begin{ruledtabular}
\setlength{\extrarowheight}{1pt}
\begin{tabular}{lcc}
\textrm{}&
\textrm{Fall}&
\textrm{Spring}\\
\colrule
Total enrolled  & 21 & 18 \\
Survey respondents  & 20 & 18 \\
Gender\\
\hspace{5mm}Women   & 20 & 18\\
 Race or ethnicity\\
\hspace{5mm}White & 9 & 9\\
\hspace{5mm}Other & 11 & 9\\
Major\\
\hspace{5mm}Biology &  12 & 11\\
\hspace{5mm}Other (Exercise Science, Dance, or & 8 & 7\\
\hspace{10mm}Health Science) &  & 
\\
Academic year\\
\hspace{5mm}Third & 14 & 13\\
\hspace{5mm}Other (Second or Fourth) & 6 & 5\\
Age (years)\\
\hspace{5mm}18--21 & 12 & 12 \\
\hspace{5mm}22 or older & 8 & 6\\
Mean (SD) final grade earned  & 90(5)\% & 86(8)\% \\
\end{tabular}
\end{ruledtabular}
\end{table}

\subsection{Data collection}

In both the fall mechanics course and spring electricity and magnetism course, the instructor administered an online survey via Qualtrics in the middle of the semester. The survey was given as a homework assignment, therefore students completed the survey outside of class. Students were awarded one percentage point on their final course grade for their completion of the survey.

The survey used existing measures of peer recognition~\cite{grunspan2016,salehi2019,bloodhart2020,sundstrom2022perceptions,sundstromWhoWhat} and peer interactions~\cite{zwolak2018educational,traxler2020network,dou2019practitioner,commeford2021characterizing,sundstrom2022interactions}, asking students to nominate peers they felt were strong in their physics course and to report peers with whom they recently interacted about their physics course. Within the recognition survey prompt, we also probed students' reasons for nominating their strong physics peers using an existing open response survey prompt~\cite{sundstromWhoWhat}: 
\begin{quote}
    \textit{Please select the students in this physics class that you think are particularly strong in the lecture material. In the text box next to each name you select, please briefly explain why you chose this student as strong in the lecture material.}\\
    \\
    \textit{Please select the students in this physics class that you had a meaningful interaction* with about the lecture material this semester.}\\
    \\
    \textit{*A meaningful interaction may mean in class, out of class, in office hours, virtually, through remote chat or discussions boards, or any other form of communication, even if you were not the main person speaking or contributing.}
\end{quote}
As mentioned in the prompts, students were provided a list of peers' names on the survey from which they selected as many names as they wanted in response to each prompt (see Appendix for a visual representation of the survey setup). Student names were listed in ascending alphabetical order by last name. For the recognition prompt, students were provided an open text box next to each peer's name to explain their nomination if they selected that peer. For the interaction prompt, ``students self-identified what counted as a meaningful interaction"~\cite[p. 6]{commeford2021characterizing}. 

In response to the recognition prompt, students selected an average of 4 and 5 peers' names in the fall and spring, respectively. In the fall, the nominations made per student for this prompt ranged from 1 to 20 and in the spring this range was 1 to 18. In both cases, one student selected almost everyone (fall) or everyone (spring) in the class, while every other student selected only a few peers. In response to the interaction prompt, students selected an average of 3 peers' names in each semester. The number of peers' names selected by each student for this prompt ranged from 1 to 7 in the fall and from 1 to 11 in the spring.

In the fall and spring semesters, respectively, 95\% and 100\% of students enrolled in the course both responded to the survey and consented to participate in research (Table~\ref{tab:demographics}). Therefore, we were able to reliably apply similar social network analysis methods to our data (more detail on these methods is provided in the next section) as in prior work~\cite{grunspan2016,salehi2019,sundstrom2022perceptions,sundstromWhoWhat} because these methods are robust to up to 30\% missing data (i.e., a response rate of at least 70\%)~\cite{smith2013structural}. For the fall semester, we included nominations made to the one non-respondent in the network diagrams (left side of Fig.~\ref{fig:networks}). We excluded this student from the remainder of the analysis because we did not have their nominations or demographic information. 

We also collected students' self-reported gender, race or
ethnicity, intended major, and academic year on the survey (Table~\ref{tab:demographics}). All survey respondents in both semesters identified as women (in the fall, the non-respondent's gender identity is unknown). In both semesters, roughly half of the students identified as White, with other students identifying as one or more of the following: American Indian or Alaska Native, Asian or Asian American, Black or African American, and Hispanic. More than half of the students in both semesters intended to major in biology, with other students intending to major in exercise science, dance, or health science. The majority of students were third-year college students, with others in their second or fourth year of college. Most students were within traditional college age (18--21 years old), while about one third of students were beyond this age.

Finally, we collected students' final course grades from the instructor at the end of each semester (Table~\ref{tab:demographics}). Students earned a mean (standard deviation) grade of 90(5)\% and 86(8)\% in the fall and spring semesters, respectively.

\subsection{Data analysis}


To address our two research questions, we conducted two stages of analysis, described in the following subsections. 
Across each stage, we drew on both social network analysis~\cite{grunspan2014,dou2019practitioner,brewe2018guide} and qualitative methods to make interpretations. Social network analysis allowed us to visualize and understand students' social positions within the class, while qualitative methods added insights to these positions. We did not aim to make any statistical claims because of the small sample size of students in the courses we investigated.

\subsubsection{Peer recognition over time}

To address our first research question, we used the survey responses to create four directed networks, one per survey prompt (peer recognition and peer interactions) and one per semester (fall and spring). In the networks, \textit{nodes} represented students and \textit{edges} represented students' selections of each other's names on the survey. Edges in both the recognition and interaction networks were considered directed in order to distinguish which student selected the other student on the survey.

We characterized the structure of each observed network to gain a baseline understanding of how students were connected to one another. We calculated the following descriptive statistics, which can be used to describe networks of any size, for each network:
\begin{itemize}
    \item \textit{Density} (measure of interconnectedness)~\cite{dou2019practitioner}: the number of observed edges relative to the maximum number of edges, calculated as 
    \begin{equation}
        \text{Density} = \frac{\text{number of observed edges}}{\text{maximum number of edges}}.
    \end{equation}
    For a directed network, the maximum number of edges is \textit{N}(\textit{N}$-$1), where \textit{N} is the number of nodes in the network.
    \item \textit{Reciprocity} (measure of the tendency for pairs of students to nominate each other)~\cite{dou2019practitioner}: the proportion of edges in the network that are two-way (i.e., student A reports a connection with student B and student B reports a connection with student A). Reciprocity is calculated as
    \begin{equation}
        \text{Reciprocity} = \frac{\text{number of two-way edges}}{\text{number of observed edges}}.
    \end{equation}
    \item \textit{Transitivity} (measure of the tendency for nodes to cluster together in triangles)~\cite{dou2019practitioner}: the proportion of two-paths (e.g., student A reports a connection with student B and student B reports a connection with student C) that close to form triangles (e.g., student A also reports a connection with student C), not considering edge direction. Transitivity is calculated as
    \begin{equation}
        \text{Transitivity} = \frac{\text{number of closed two-paths}}{\text{number of two-paths}}.
    \end{equation}
    \item \textit{Indegree centralization} (measure of skewness in the distribution of incoming edges)~\cite{butts2006exact}: the sum of differences in \textit{indegree} (the number of other students that selected a given student) between the node with the highest indegree (receiving the most nominations) and every other node in the network, divided by the maximum possible sum of differences of indegree for all nodes. This maximum is (\textit{N}$-$1)$^2$, which would arise if the network were a perfect ``star" in which only one node is connected to any of the other nodes and that one node has only an incoming edge from every other node. Indegree centralization values range from 0 and 1 with values closer to 1 indicating that nominations are highly concentrated around a small subset of students in the network. For a directed network with \textit{N} nodes, individual nodes \textit{$v_i$} with indegree \textit{$c(v_i)$}, and one node \textit{$v_*$} with the highest indegree \textit{$c(v_*)$}, indegree centralization is calculated as
    \begin{equation}
        \text{Indegree centralization} = \frac{\sum_{i=1}^{N} (c(v_*)-c(v_i))}{(N-1)^2}.
    \end{equation}
\end{itemize}
We reflect on the raw values of these statistics when we describe the results of the study. However, we did not calculate the standard errors of the statistics because we did not aim to make any statistical comparisons between values.


\begin{table*}[t]
\caption{\label{tab:codingscheme}%
Definitions of codes from Ref.~\cite{sundstromWhoWhat} applied in this study and example responses from this study.}
\begin{ruledtabular}
\setlength{\extrarowheight}{1.2pt}
\begin{tabular}{p{0.12\linewidth}  p{0.4\linewidth} p{0.36\linewidth}}
Code & Definition & Example\\
\hline
Participation & \hangindent=1em Active contributor to in-class discussions and activities & \hangindent=1em“She openly asks questions and answers the questions.”\\
Understanding & Knowledgeable about the course material & \hangindent=1em``Seems to have a strong sense of the topics in class." \\
Performance& \hangindent=1em Receiving good grades; answering questions correctly & “Her test scores reflect that she is strong.” \\
Problem solving & \hangindent=1em Visualizing or reasoning through problems; applying the right equations to problems  & \hangindent=1em ``They apply the lecture material in lab and group problems." \\
Motivation & \hangindent=1em Putting a lot of time or effort into the course; determined &  \hangindent=1em “Watches lecture videos before class and takes notes to be familiarized with the concepts.”\\
Helping & \hangindent=1em Providing support with the course material to others (nominator mentions benefitting from this support) &  \hangindent=1em“I regularly work with [her] and whenever I have questions she is able to guide me to the method to solve the problem.”
\\
Natural ability & \hangindent=1em Having an innate aptitude for understanding the course material; understanding content quickly &  \hangindent=1em “She seems to understand the concepts quickly.”\\
Explaining & \hangindent=1em Describing or clarifying the course material to others &  ``Explains with clarity."\\
Other &  \hangindent=1em Part or all of the explanation is vague or does not fit with above codes & ``Because she sounds confident in her answers." \\
None & No explanation provided & [Blank] \\
\end{tabular}
\end{ruledtabular}
\end{table*}


Similar to prior work~\cite{grunspan2016,salehi2019}, we also characterized the recognition celebrities by examining the top nominees of each recognition network. We examined four celebrities per network because the number of nominations received by each student dropped to five or fewer after the fourth most nominated student (who received 8 and 7 nominations in the fall and spring, respectively), and at least three students received five nominations in each network. Including the students receiving five nominations would consider almost half of the class as celebrities, and we wanted the celebrities to only include the most renowned students. 


We then examined whether and how the individual celebrities (top four nominees) in the recognition networks changed between the fall and spring semesters. To examine recognition over time for the whole class, we compared students' recognition network indegree (number of received nominations as strong in the course) in the fall to their recognition network indegree in the spring by qualitatively examining a scatterplot of these two variables. This full-class analysis only included the 18 students who were enrolled in both the fall and spring courses.

\subsubsection{Student reasons for nominating strong physics peers over time}

To address our second research question, we first determined whether and how students' nominations of strong peers related to student outspokenness. We analyzed the open response survey prompt asking students to explain why they nominated their peers as strong in the course. The first author read all of the responses to get a sense of the data as a whole~\cite{Tesch1990} and determined that students' ideas were similar to those in our previous study which analyzed the same survey prompt for students at a different institution~\cite{sundstromWhoWhat}. Therefore, we applied the coding scheme established in our prior work~\cite{sundstromWhoWhat} which included verbal participation, i.e., outspokenness, as one of the skills for which students recognized their strong physics peers. The first author achieved sufficient interrater reliability with two other coders for this coding scheme in our prior study, so this author coded all of the responses in the current study. Only a subset of the original coding scheme was applied to the current data set, however, likely due to differences in instructional style and student population between the two studies. For example, we did not apply the \textit{Experience} code~\cite{sundstromWhoWhat}, which captures ideas such as having taken an AP physics course or having work experience relevant to the course, in this study because no students in the current data set were pursuing a major in physics. Table~\ref{tab:codingscheme} lists the subset of codes and definitions we applied in the current study as well as example responses from this data set.

With the coded responses, we examined the frequencies of all of the codes in the fall and spring semesters separately to determine the most prevalent skills for which students recognized their strong peers and whether and how the most frequent codes changed across the course sequence. We paid particular attention to the \textit{Participation} code, which serves as a measure of outspokenness during class. In this analysis, we excluded responses for which students selected a peer to nominate but left the explanation of their nomination blank (\textit{None} code). Of the 82 and 60 total responses in the fall and spring semesters, respectively, 46 and 29 of these responses were not left blank and were therefore included in this part of the analysis (all survey selections, or edges in the networks, are included in the rest of our analysis). We also looked at the frequencies of codes used to describe each of the individual celebrities. 


We then investigated the relationship between students' total degree in the interaction network (sum of their reported interactions and the number of other students who reported interacting with them) and indegree in the recognition network (number of received nominations as strong in the course) for each semester. We qualitatively examined a scatterplot of these two variables for all students to understand the extent to which students who were centrally positioned in the interaction network, and therefore outspoken in direct peer interactions, received recognition from peers as strong in the course. We used our findings in conjunction with the explanations analysis described above to disentangle the roles of different types outspokenness in peer recognition. If, for instance, in-class \textit{Participation} was prominent in the written explanations but there was little association between students' interaction degree and received recognition, then we might infer that student outspokenness in front of the whole class or with the instructor were more important for gaining recognition than having many peer interactions.

In this comparison of interactions and recognition, we used total degree rather than indegree to characterize students' position in the interaction network because interactions are inherently two-way and counting both incoming and outgoing edges balances possible biases from under-reporting due to recall bias (e.g., student A reporting an interaction with student B but student B not reporting an interaction with student A because they forgot student A's name) and over-reporting (e.g., student A reporting an interaction with student B but student B not reporting an interaction with student A because they did not consider the interaction meaningful)~\cite{sundstrom2022interactions}.


We also investigated the extent to which students' received recognition from their peers related to their course performance by first looking at the frequencies of the \textit{Performance} code in students' explanations of their nominations. We then compared students' recognition network indegree (number of received nominations as strong in the course) to their final course grade in each semester. We report qualitative patterns from a scatterplot of these two variables.

\begin{figure*}[ht!]
    \centering
    \includegraphics[width=6.7in,trim = {0 0 5cm 0}]{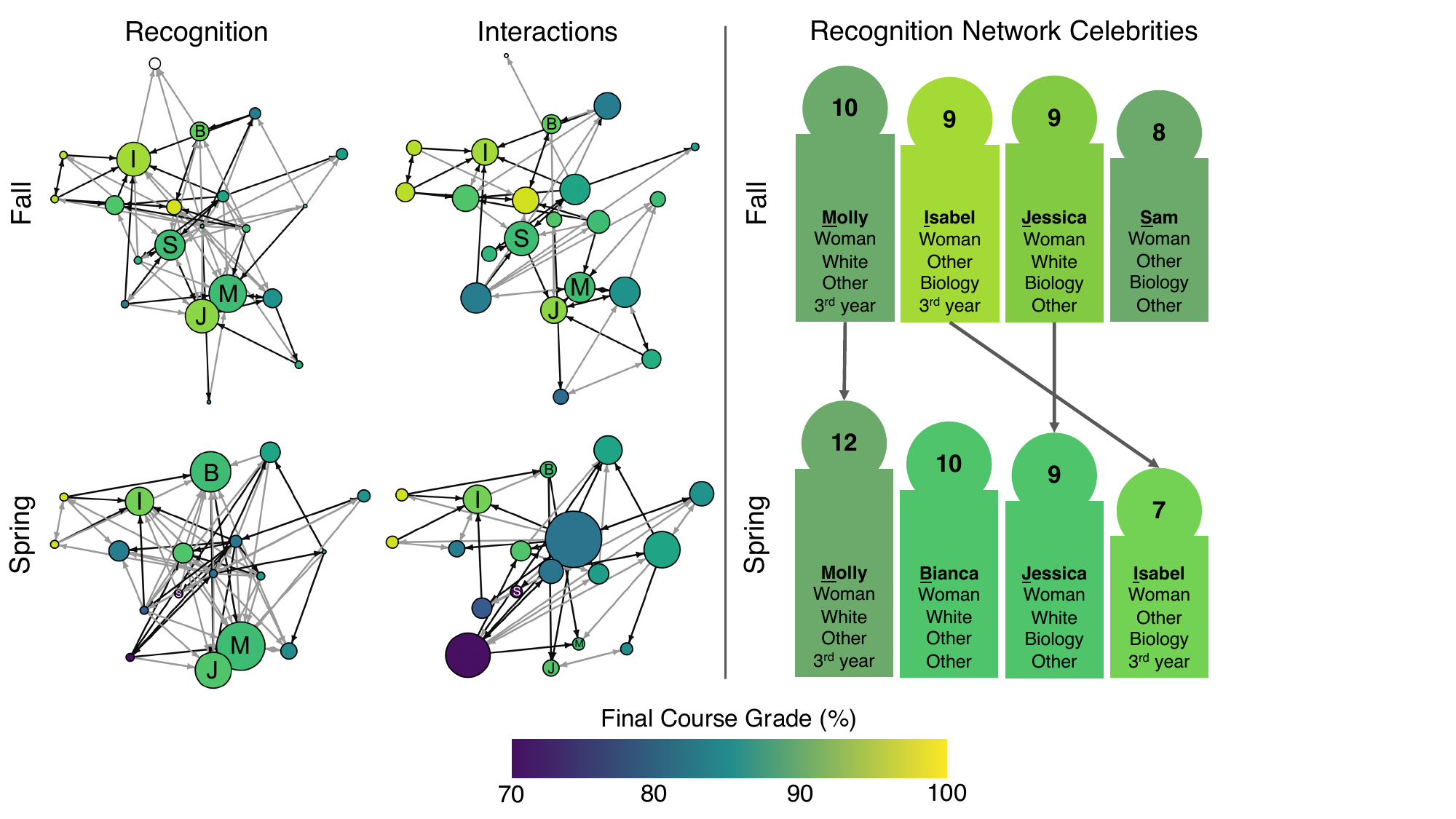}
    \caption{On the left are the network diagrams for all four observed networks. Nodes represent students and are located in the same position in every network, using a spring-based layout for the fall recognition network as a basis for generating the node locations (there are three more nodes, located on the perimeter, in the fall networks than in the spring networks; see Table~\ref{tab:demographics}). Node labels indicate celebrities, marked by the first letter of their name (shown on the right). Node color indicates students' final course grade in a given semester, with white representing an unknown grade for the one non-respondent in the fall. In the recognition networks, nodes are sized proportional to indegree (number of received nominations as strong in the course material) and in the interaction networks, nodes are sized proportional to total degree (total number of adjacent edges). Edges point from the nominator to the nominee. Black edges indicate edges that occur in both a recognition network and its corresponding interaction network (and vice versa) and gray edges indicate edges in a recognition network that do not occur in its corresponding interaction network (and vice versa). On the right are the celebrities--students receiving the highest number of nominations--of each recognition network. Numbers in the circles indicate the number of nominations received and colors indicate final course grade. Text boxes show students' names (all are pseudonyms) and self-reported gender, race or ethnicity, major, and academic year. Arrows show the trajectories of students who are celebrities in both semesters. 
    }
    \label{fig:networks}
\end{figure*}

\section{Results}

We present the results in order of our two research questions.

\subsection{Peer recognition over time}

The left side of Fig.~\ref{fig:networks} and Table~\ref{tab:networkstats} show the network diagrams and network statistics, respectively, for the four observed networks. We cannot directly compare densities between the fall and spring networks in this study, or compare the densities of the networks in this study to those in previous studies, because density does not scale linearly with network size. However, all four networks are relatively dense, containing many edges that connect the nodes. There are no isolated nodes, nodes with zero adjacent edges indicating students with zero recognition or interaction edges, in any network. Additionally, every node has at least one incoming edge in every network, indicating that every student was selected by at least one peer on both the recognition and interaction survey prompts in each semester. This interconnectedness is also reflected in the fairly high transitivity in all four networks, indicating a tendency for nodes to form small, connected clusters.  

\begin{table}[t] 
\centering
\caption{\label{tab:networkstats}
Descriptive statistics for the four observed networks. 
}
\begin{ruledtabular}
\setlength{\extrarowheight}{1pt}
\begin{tabular}{lcccc}
 & \multicolumn{2}{c}{Recognition} &  \multicolumn{2}{c}{Interactions} \\ 
 \cline{2-3}
 \cline{4-5}
 & Fall & Spring  & Fall  & Spring \\ 
 \hline
 Nodes & 21 & 18 & 21 & 18 \\ 
 Edges & 82 & 78 & 60 & 51 \\ 
 Density & 0.21 & 0.25 & 0.16 & 0.17 \\ 
 Reciprocity &  0.25 & 0.23 & 0.51 & 0.47\\ 
 Transitivity &  0.43 & 0.53 & 0.29 & 0.40 \\ 
 Indegree centralization & 0.34 & 0.45 & 0.17 & 0.13 \\
\end{tabular} 
\end{ruledtabular}
\end{table} 

We observe higher reciprocity, the fraction of edges that are two-way rather than one-way, in the interaction networks than in the corresponding recognition networks. Relatedly, the indegree centralization values suggest that the incoming edges (indicating the students who are selected by others on the survey) in the recognition networks are much more concentrated around one or a few students than those in the corresponding interaction networks. These structural differences are unsurprising due to the mutual nature of interactions, in which two students are both engaged in a conversation with one another, whereas peer recognition is more often unidirectional: one student recognizes their peer as a strong physics student, but their peer does not recognize them as a strong physics student.

The right side of Fig.~\ref{fig:networks} characterizes the individual celebrities in each recognition network, determined as the four students receiving the most nominations from their peers as strong in their physics course. In the fall, the top nominated students are Molly, Isabel, Jessica, and Sam, respectively (all student names are pseudonyms). Molly identifies as a White woman and is a third year student studying a non-biology major. Isabel identifies as a non-White woman and is a third year student studying biology. Jessica identifies as a White woman and is pursuing a biology major. Sam identifies as a non-White woman studying biology. In the spring, Molly, Isabel, and Jessica remain celebrities, while Sam does not. The fourth celebrity in the spring is Bianca, who identifies as a White woman and is studying a non-biology major.

\begin{figure}[t]
    \centering
    \includegraphics[width=3.4in,trim={0 5cm 12cm 0}]{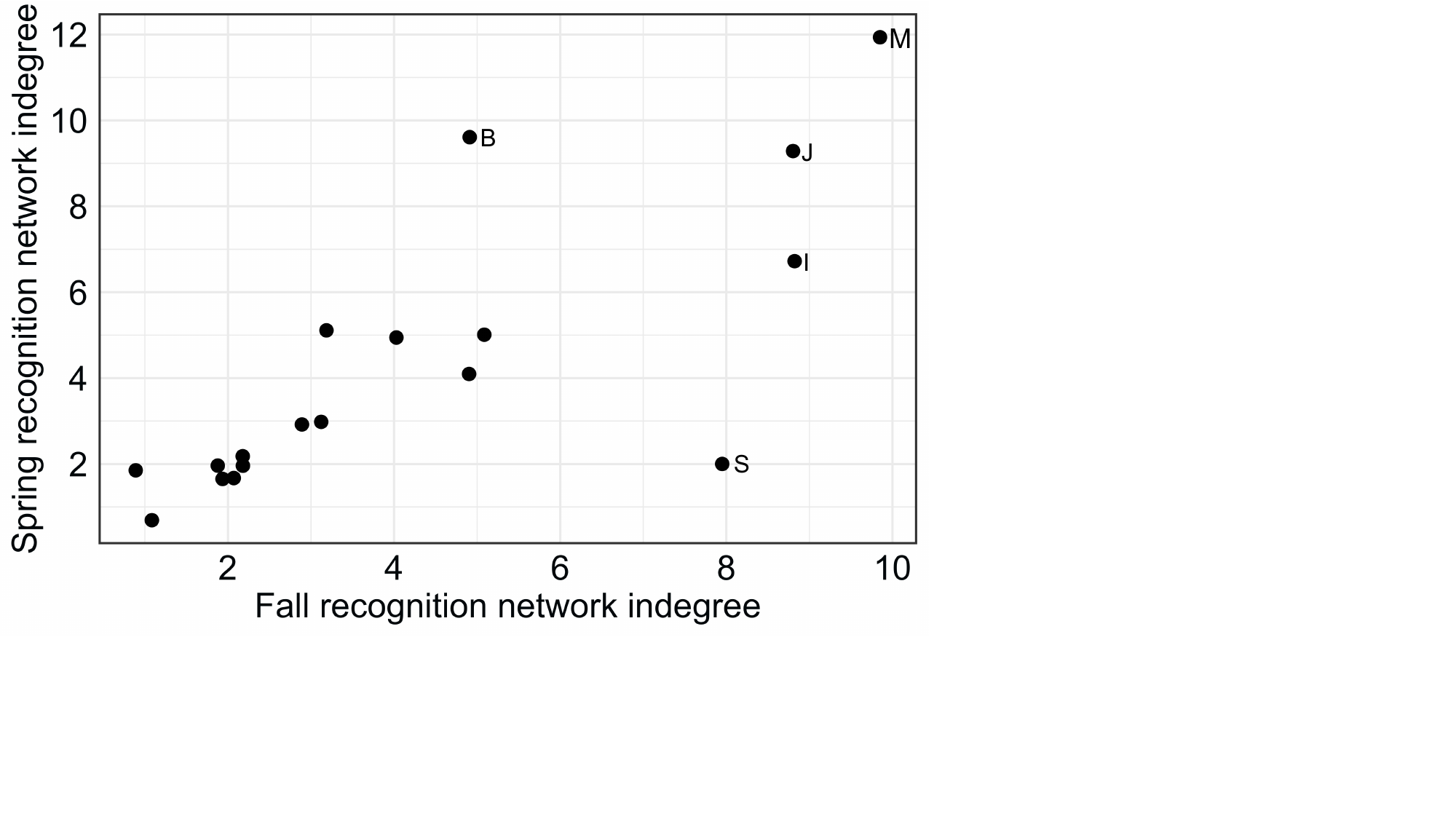}
    \caption{Scatterplot comparing students' recognition network indegree (number of nominations received as strong in the course material) in each semester for the 18 students who were enrolled in both semesters. All values are integers but points are jittered to separate overlapping points. Labels indicate celebrities, marked by the first letter of their name (shown in Fig.~\ref{fig:networks}). The plot indicates a positive trend between fall and spring recognition, with two notable exceptions: Sam and Bianca.
    }
    \label{fig:semesterscatter}
\end{figure}

We see some evidence that peer recognition may be stable over time. Three out of the four recognition celebrities in the fall (Molly, Isabel, and Jessica), for example, are also celebrities in the spring (right side of Fig.~\ref{fig:networks}). Examining the whole class, we similarly observe that students who receive more nominations as strong in the course in the fall also tend to receive more nominations in the spring (upward trend in Fig.~\ref{fig:semesterscatter}). Correspondingly, in the recognition network diagrams (left side of Fig.~\ref{fig:networks}), the size of each node (representing indegree) in the fall recognition network appears to be correlated with the size of the corresponding node in the spring recognition network (i.e., nodes that are larger in the fall also tend to be larger in the spring).  

There are, however, two substantial exceptions to this pattern: Sam receives eight nominations in the fall (when she is a celebrity) and only two nominations in the spring (when she is not a celebrity) and Bianca receives five nominations in the fall (when she is not a celebrity) and 10 nominations in the spring (when she is a celebrity; Fig.~\ref{fig:semesterscatter}). Thus, despite the overarching pattern that recognition seems to be fairly stable over time, the cases of Sam and Bianca indicate that the students receiving the most peer recognition--the celebrities--are subject to fluctuations over time. Even in this small class with a nearly identical set of students enrolled throughout the course sequence, we observe that individuals' celebrity status can change dramatically and in either direction (i.e., losing celebrity status over time or gaining celebrity status over time). In the following two subsections, we add nuance to this finding by analyzing the reasons students give for nominating their peers throughout the course sequence.

\subsection{Student reasons for nominating strong physics peers over time}

\subsubsection{Peer recognition and outspokenness}

\begin{figure}[t]
    \centering
    \includegraphics[width=2.8in,trim={0 0 18cm 0}]{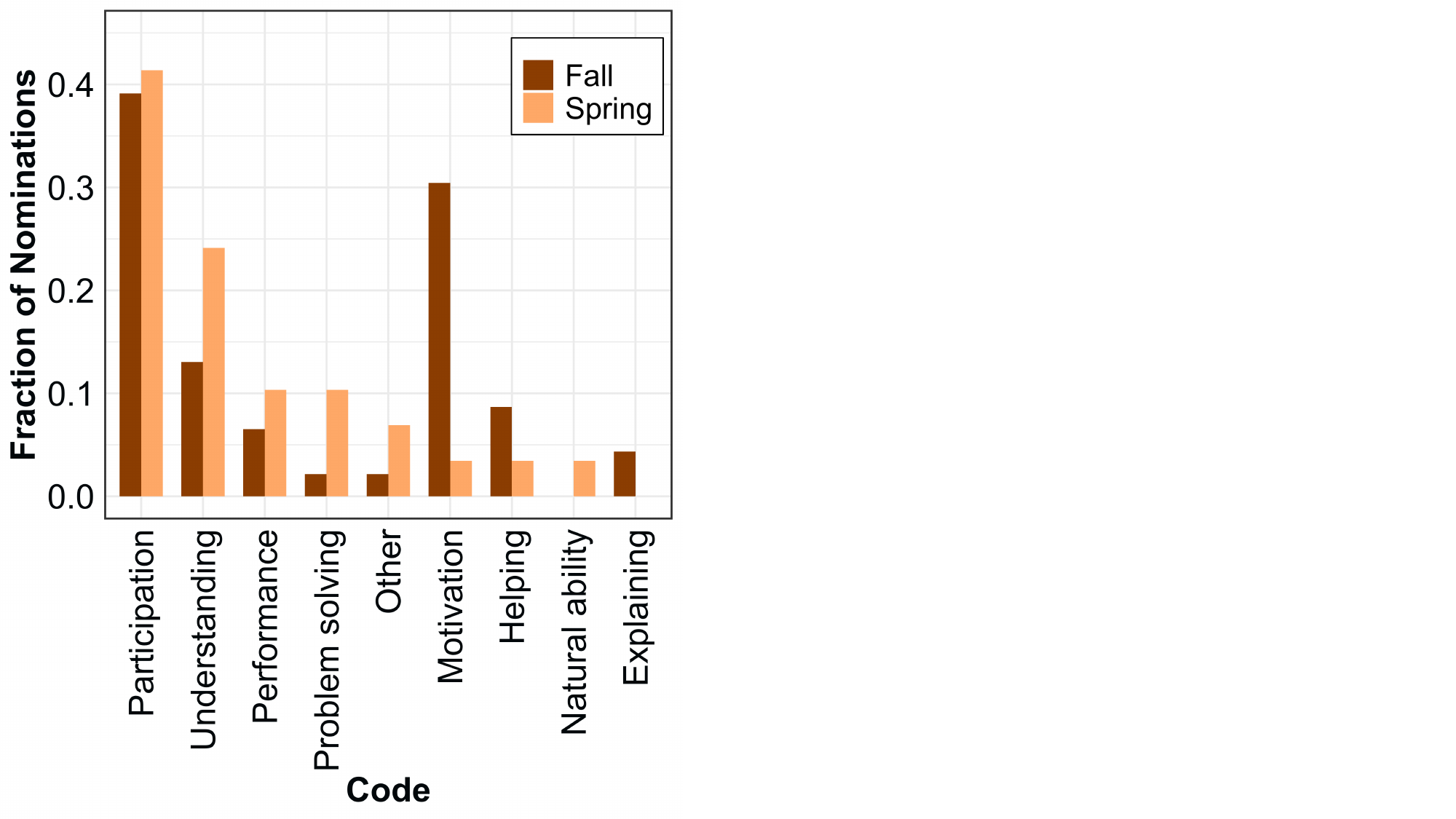}
    \caption{Fraction of all non-blank explanations in each recognition network containing each code from the coding scheme. Fractions do not necessarily sum to one because each explanation could receive multiple codes, though on average each explanation received 1.05 codes. We coded 46 and 29 non-blank explanations in the fall and spring, respectively.
    }
    \label{fig:codes}
\end{figure}

Our analysis of students' written explanations for why they nominated their strong peers suggests that the skills for which students are recognized change slightly over time (Fig.~\ref{fig:codes}). In the fall (brown bars), students most frequently describe strong peers as those who verbally participate, are motivated, and exhibit understanding. In the spring (orange bars), on the other hand, student explanations are mostly about verbal participation and understanding, and they are no longer about motivation. Overall, in-class outspokenness is the most prominent reason for which students recognize their strong physics peers, with about 40\% of explanations receiving the \textit{Participation} code in each semester.

\begin{figure}[t]
    \centering
    \includegraphics[width=3.1in,trim={0 0 25cm 0}]{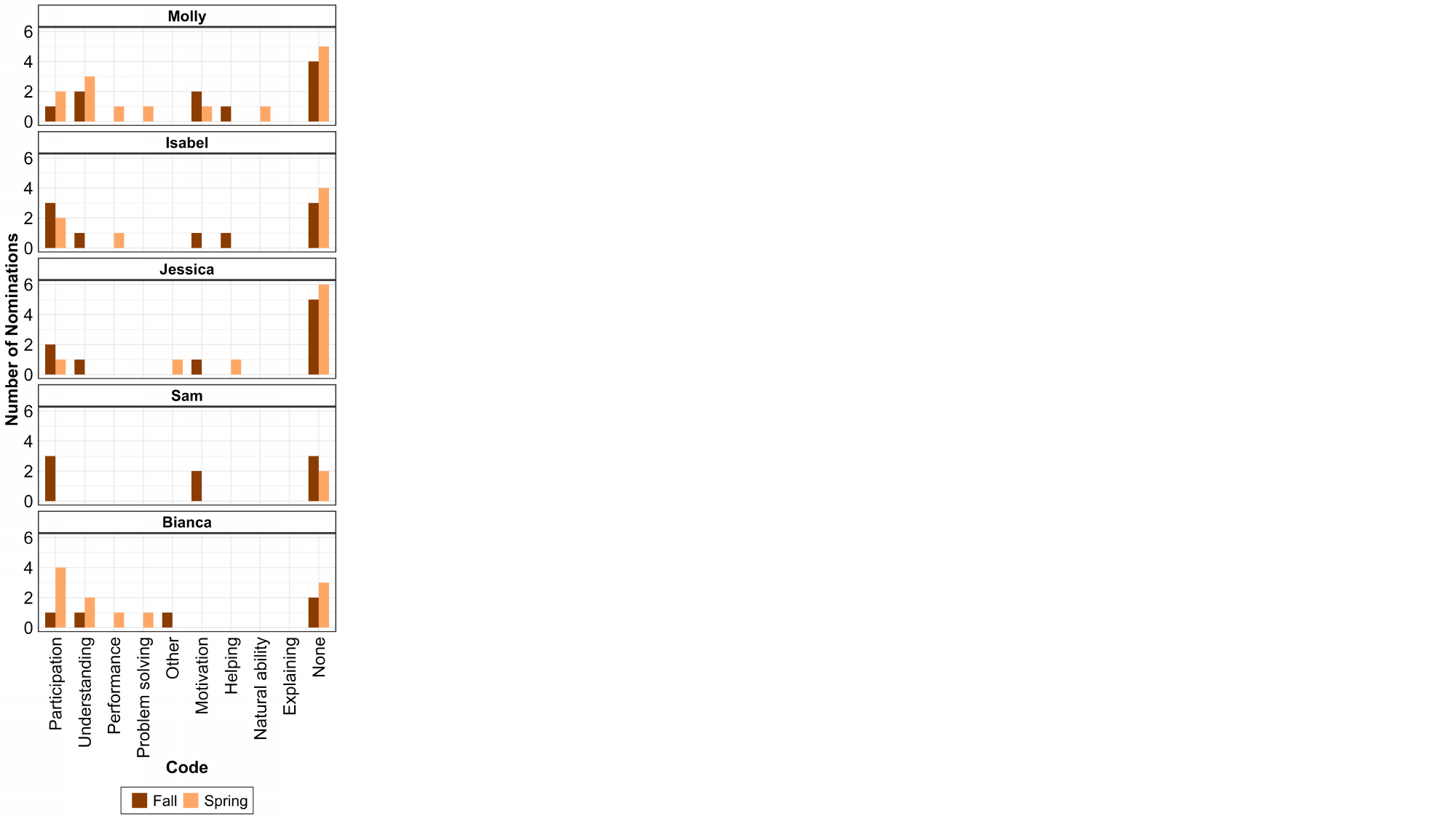}
    \caption{Number of explanations containing each code from the coding scheme in student nominations of the celebrities identified in Fig.~\ref{fig:networks}. Numbers for a given celebrity in a given semester do not necessarily sum to their total number of received nominations because each explanation could receive multiple codes, though on average each explanation for a celebrity received 1.04 codes.
    }
    \label{fig:celebcodes}
\end{figure}

These patterns in student explanations of their nominations across the whole class are mostly consistent with the patterns in student explanations of the individual celebrities (Fig.~\ref{fig:celebcodes}). The most prominent codes in each semester are spread across the celebrities, and the celebrities in both semesters are all described by at least one peer as being outspoken. 

It also seems that Sam and Bianca's recognition trajectories directly track with their outspokenness: each of them is recognized by multiple peers for being outspoken during the semester in which they are a celebrity, whereas they each receive far fewer nominations for being outspoken in the semester when they are not a celebrity (Fig.~\ref{fig:celebcodes}). Anecdotal information from the course instructor allows us to better understand Sam and Bianca's situations. In the fall mechanics course, Sam attends most class sessions and actively participates in class, and she is a recognition celebrity in this course. Due to a family situation, Sam only attends every other class session in the spring semester and the days that she does attend, she sits in the very back to catch up on work and does not participate much. With limited visibility and less participation in class, Sam receives much less peer recognition in the spring despite being surrounded by the same classmates who highly recognized her in the fall. Bianca, on the other hand, has a conflicting class in the fall semester and so only attends the first or second half of each lab and GP session. She is not a recognition celebrity in the fall, likely because of this limited exposure in front of her peers and because she does not speak up much during class (possibly due to a lack of confidence in her grasp of the course material given her attendance). In the spring, however, Bianca attends most of the class sessions and frequently answers questions during lecture, and she becomes a recognition celebrity.

\begin{figure}[t]
    \centering
    \includegraphics[width=3.4in,trim={0 5cm 12cm 0}]{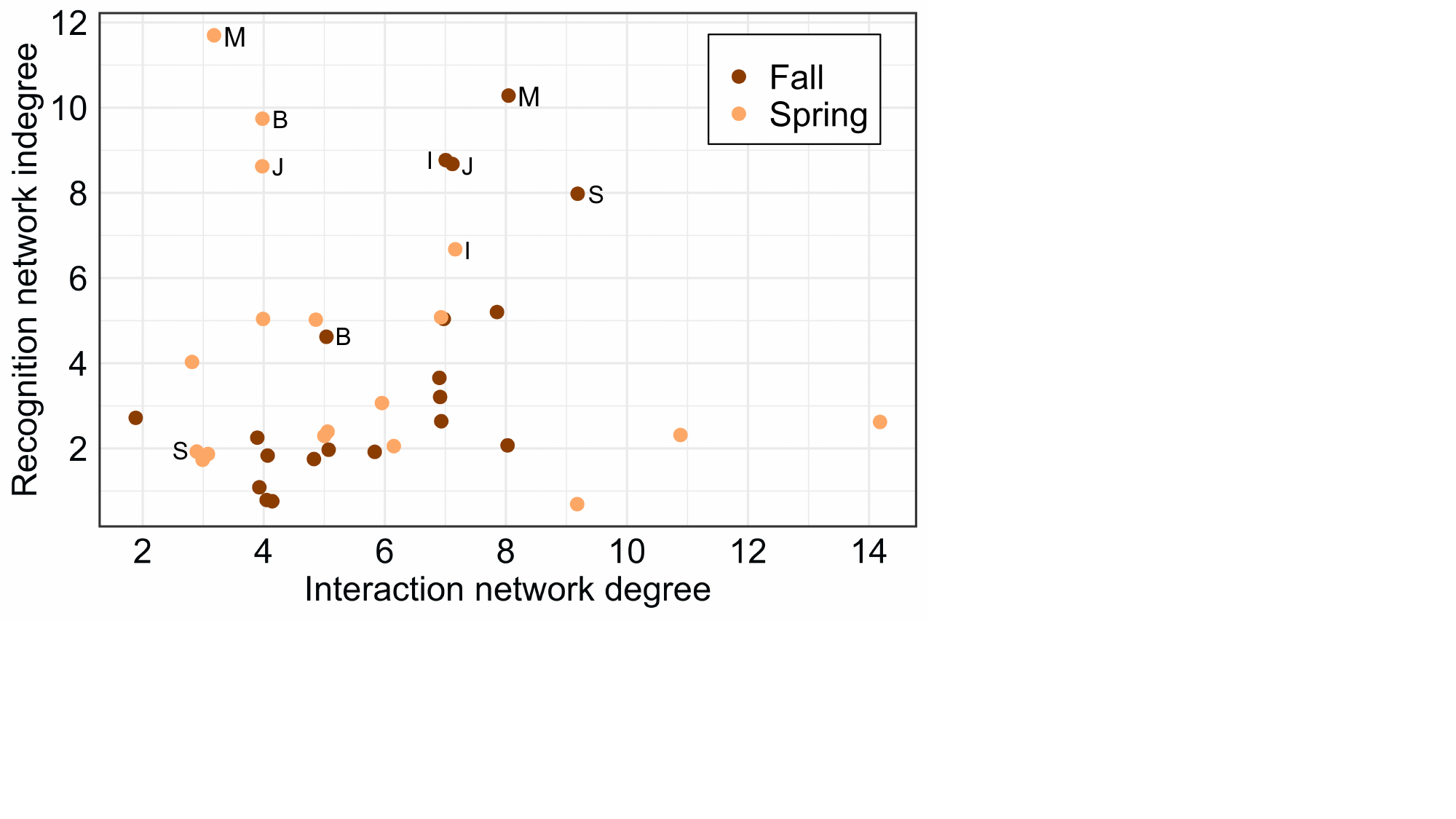}
    \caption{Scatterplot comparing students' interaction network indegree (total number of connections) to their recognition network indegree (number of nominations received). Colors indicate semester. All values are integers but points are jittered to separate overlapping points. Labels indicate celebrities, marked by the first letter of their name (shown in Fig.~\ref{fig:networks}). The plot indicates a positive trend in the fall semester, but no trend in the spring semester.
    }
    \label{fig:interactionscatter}
\end{figure}

In summary, the cases of Sam and Bianca, in addition to our analysis of all students' reasons for nominating peers, highlight the consistently strong association between being outspoken and receiving peer recognition. We build on this observation by also determining what kind of outspokenness is related to receiving peer recognition--whether through peer interactions or other means. We see an upward trend between students' interaction network degree and recognition network indegree in the fall (brown dots), but no clear trend between these variables in the spring (orange dots; Fig.~\ref{fig:interactionscatter}). Moreover, in the fall, all four recognition celebrities have relatively  high interaction network degrees. In contrast, there are a few students with much higher interaction network degrees than the recognition celebrities in the spring. We can also see these relationships in the network diagrams (left side of Fig.~\ref{fig:networks}): there appears to be a correlation between node size in the recognition network (indicating indegree) and node size in the interaction network (indicating total degree) in the fall, but not in the spring. There are large dark blue and purple nodes in the spring interaction network, for example, that are quite small in the spring recognition network, representing students who frequently interact with their peers but whom do not receive much peer recognition.

These results suggest that being outspoken in peer interactions is important for receiving peer recognition in the fall course, but not the spring course. Other forms of outspokenness during class, such as answering questions in front of the whole class, are likely more important for receiving peer recognition in the spring.

\subsubsection{Peer recognition and academic performance}

\begin{figure}[t]
    \centering
    \includegraphics[width=3.4in,trim={0 5cm 12cm 0}]{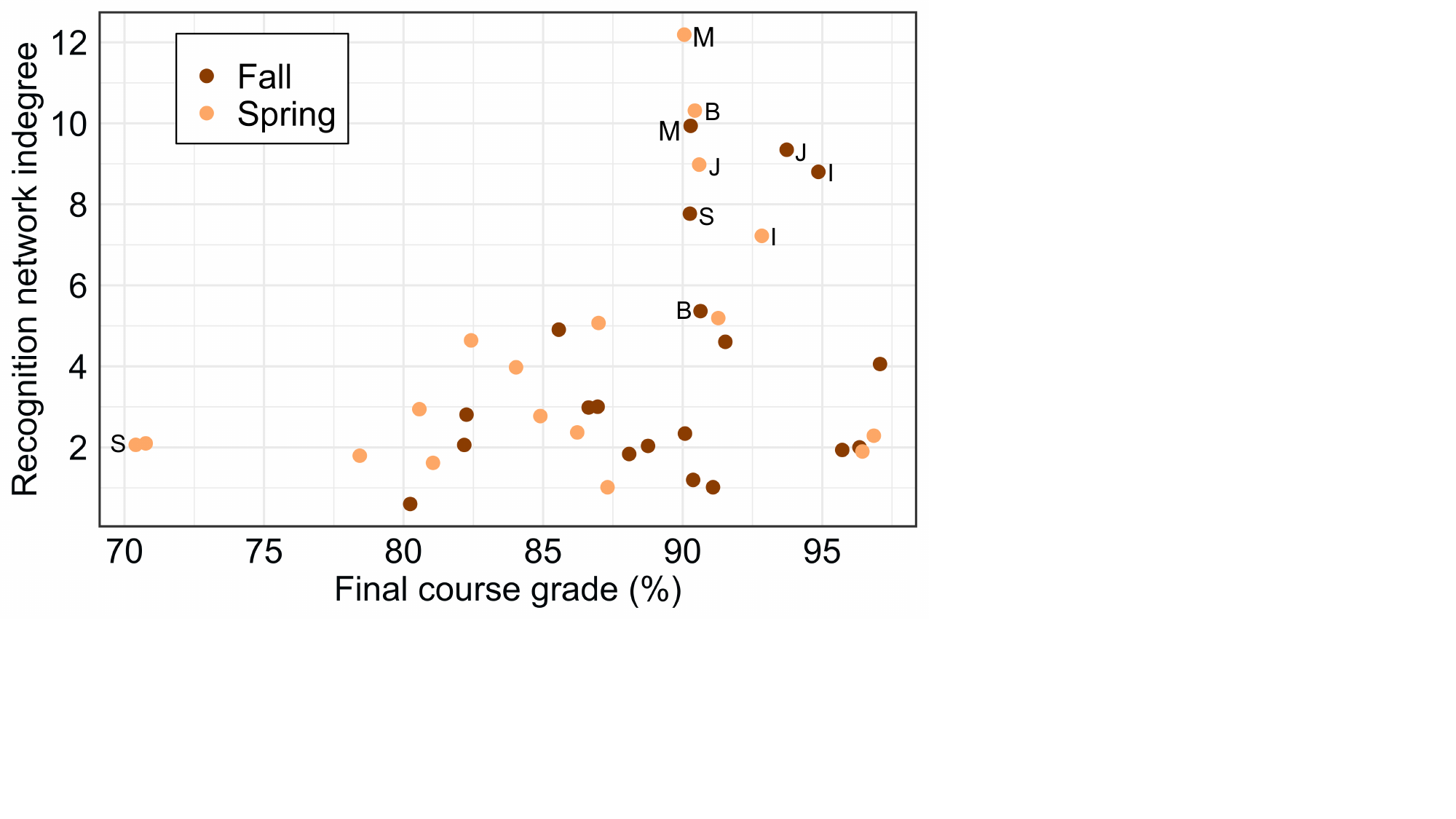}
    \caption{Scatterplot comparing students' final course grade (as a percentage) to their recognition network indegree (number of nominations received as strong in the course material). Colors indicate semester. All indegree values are integers but points are jittered to separate overlapping points. Labels indicate celebrities, marked by the first letter of their name (shown in Fig.~\ref{fig:networks}). The plot shows no strong trend in either semester.
    }
    \label{fig:gradescatter}
\end{figure}

Students do not frequently describe their strong peers as having high performance in the class, with 10\% or fewer of student explanations mentioning academic achievement in each semester (\textit{Performance} code in Figs.~\ref{fig:codes} and~\ref{fig:celebcodes}). Similarly, we see no strong association between students' final course grade and their recognition network indegree in either the fall or the spring semester (Fig.~\ref{fig:gradescatter}). Indeed, we observe that in the fall (brown dots) and spring (orange dots), respectively, there are three and two students earning higher final course grades than the four celebrities, but these students do not receive substantial peer recognition (bottom right of Fig.~\ref{fig:gradescatter}, indicating high final course grade and low recognition network indegree). 

This relationship between peer recognition and student performance is also evident in the recognition network diagrams (left side of Fig.~\ref{fig:networks}). Within each recognition network, the color of the nodes (representing final course grade) is not strongly correlated with the size of the nodes (representing recognition network indegree). For example, there are both large and small nodes with light green or yellow coloring in the fall recognition network, indicating that students who receive high final course grades receive a wide range of recognition from their peers. 

At the same time, however, all recognition celebrities seem to meet some threshold level of performance. In each semester, the students receiving the most nominations as strong in the course earn a grade of at least 90\% (top right of Fig.~\ref{fig:gradescatter}, indicating high recognition network indegree and high final course grade). Indeed, the largest nodes in the recognition network diagrams, indicating the recognition celebrities, are all medium to light green in color, indicating that the most recognized students earn a final course grade of at least 90\% (left side of Fig.~\ref{fig:networks}).

Earning a high grade is therefore necessary, but not sufficient, for becoming a recognition celebrity. Given our results related to outspokenness discussed in the previous section, it seems that both earning a high grade \textit{and} being outspoken are necessary to receive a significant amount of peer recognition.

\section{Discussion}

In this study, we uniquely examined peer recognition for a single cohort of students across a two-semester introductory physics sequence at a mostly-women, liberal arts institution. We contribute to the existing research literature by largely isolating the effect of time on students' nominations, and reasons for these nominations, of strong physics peers in order to test prior work's hypothesis that peer recognition changes over the time-scale of semesters~\cite{grunspan2016,salehi2019,bloodhart2020,sundstrom2022perceptions,sundstromWhoWhat}.

\subsection{Changes in who gets recognized over time}

For most of the students in our analysis, the extent to which they were recognized by their peers did not vary much between semesters, consistent with prior work showcasing the stability of first impressions~\cite{grunspan2016,buchert2008first,laws2010student}. We also found that three out of the four recognition celebrities in the fall were also celebrities in the spring.

At the same time, however, we observed drastic changes in the extent to which two students (Sam and Bianca) were recognized by their peers. Each of these students were highly recognized in one semester, but were not recognition celebrities in the other semester. These shifts in peer recognition occurred despite the rest of the students enrolled, and the instructor and instructional style, in the two courses remaining nearly identical. 

Such findings provide nuanced support to prior work's hypothesis that peer recognition changes over the time-scale of semesters~\cite{grunspan2016,salehi2019,bloodhart2020,sundstrom2022perceptions,sundstromWhoWhat}. It seems that while peer recognition can change across a course sequence, patterns of recognition--such as there being a gender bias in first-year, but not beyond first-year, science courses--may not be as universal as previously thought. Indeed, our results challenge the generalizability of previous findings that gender bias does not persist beyond first-year science courses: peer recognition appears to be course-specific and highly sensitive to variables such as fluctuations in which students most strongly participate in the class. We recommend for future work to better understand how patterns of peer recognition arise in a given course setting, such as by more closely investigating how the social dynamics within or characteristics of a given course translate to patterns of peer recognition. For example, researchers may conduct classroom observations and/or student interviews along with collecting survey data about peer recognition. 

As for instructional implications, if patterns in peer recognition--such as gender biases--are unlikely to follow a consistent pattern over time, interventions aimed at equitable peer recognition may need to be implemented consistently throughout a course sequence, rather than just early on in the course sequence as prior work suggests~\cite{sundstrom2022perceptions}. We discuss what such interventions may look like in the next section.

\subsection{Changes in what gets recognized over time}

Our analysis relating peer recognition to both student outspokenness and academic performance illuminates how the reasons for which students recognize their strong peers change over time.

We found across both semesters that outspokenness is the most frequent skill for which students recognize their strong physics peers and that the top four recognition celebrities were all considered outspoken. In other words, outspokenness is consistently correlated with getting recognized, similar to prior work in this educational context~\cite{grunspan2016,sundstrom2022perceptions}. As observed here (i.e., with the cases of Sam and Bianca), however, even among the same set of students, the degree to which an individual is outspoken can change over time and correspond to changes in who gets recognized.


We expanded on prior work by also identifying the specific kinds of outspokenness that are related to peer recognition over time. Results indicate that outspokenness in direct peer interactions is associated with receiving peer recognition in the fall, but not in the spring. Other forms of outspokenness in class are likely more important for recognition in later courses, such as asking and answering questions during whole-class discussions or frequently talking to the instructor during group activities in front of peers.


This distinction is important for determining appropriate instructional interventions, for instance to mitigate gender bias in peer recognition. Previous research has found that men are often more outspoken than women in front of the whole class in science courses~\cite{eddy2014,eddy2015caution,aguillon2020gender,nichols2022participation}. Our findings suggest that both during \textit{and} beyond the first course in a sequence (when students seemingly get to know each other's strengths through peer interactions), it is important for instructors to consistently offer opportunities for a diverse set of students to participate in whole-class discussions. For example, instructors may use ``warm'' or cold call when choosing students to present ideas~\cite{tanner2013structure} so that outspokenness is distributed across many students rather than only a few. Additionally, during small group activities in class, instructors may allot roughly even amounts of time with each group so that all students have the opportunity to discuss content and ask questions to the instructor in front of their groupmates. Based on our study, these practices are likely more important than diversifying peer interaction networks due to the diminishing association between peer interactions and recognition over time.


Finally, our results relating peer recognition to academic performance are consistent with prior work demonstrating that final course grade is not sufficient for being a recognition celebrity and that some high performers are not celebrities~\cite{grunspan2016,salehi2019}. This could mean that, in both courses in the sequence, students did not know the celebrities' course grades and/or did not use those grades as a basis for nominating them--they may have developed their perceptions of the celebrities in other ways. However, we found that both earning a high grade \textit{and} being outspoken is important for receiving significant peer recognition. It could be the case that outspokenness and performance are correlated and performance merely serves as a proxy for participation, which seems to be a strong basis for recognition. Students who are doing well in the course may be more confident to, for example, raise their hand during lectures and answer questions in front of the class. In our study, the data point of Sam provides preliminary evidence of this possibility: she is both outspoken and high performing in the fall, and neither outspoken nor high performing in the spring (Figs.~\ref{fig:celebcodes} and~\ref{fig:gradescatter}). Future work should continue to disentangle the effects of grades and outspokenness on peer recognition.

\subsection{Limitations}

With regard to the network survey, we may not have captured all recognition nominations and peer interactions due to recall bias, where students may forget their peers' names or forget interactions with their peers. Students also filled the survey out as part of a homework assignment outside of class when they were not surrounded by their physics peers. We believe our data still captures a fairly accurate picture of peer recognition and peer interactions in these courses because we provided a full class roster on the survey to assist with students remembering their peers' names. Future work, however, should directly compare the effects of students filling out the survey in class versus outside of class and using free recall versus selection from a class roster on student responses.

Second, this study took place with a small number of students at one institution. We used multiple forms of analysis to triangulate the data as much as possible to account for this small sample size in our interpretations. Nevertheless, we suggest for researchers to collect more data from similar instructional contexts in order to make more generalizable claims.

In addition, we were able to control for many variables between the semesters we analyzed: instructor, instructional style, student enrollment, student gender, and student academic year. However, we could not control for course content (i.e., mechanics versus electricity and magnetism) as it is subject to change between semesters. While we do not have data about students' incoming knowledge available for the student participants in this study, it is possible, for example, that some students may have had prior experience with mechanics but not with electricity and magnetism (e.g., from high school) and then received more recognition in the fall mechanics course than the spring electricity and magnetism course. We found that most students received similar amounts of recognition in both semesters (with the exceptions of Sam and Bianca whose personal situations explain their differing recognition between semesters), however, limiting the possibility that our results related to time are impacted by the changing course content. Moreover, students' prior knowledge is likely reflected in their grades, which we observed to be necessary, but not sufficient, for gaining recognition. Still, future research should more closely investigate the effects of students' incoming knowledge about the course content on peer recognition.

Finally, we studied a cohort of students across a two-semester course sequence. While this expands the time-scale of prior work on peer recognition, we could not follow these students any longer because these were the only two physics courses offered at the institution and the university did not offer a physics major. We recommend for future work to conduct a similar analysis over an even longer time-scale, such as by following cohorts of physics students throughout their entire undergraduate program.

\vspace{0.4cm}

\section{Conclusion}

Researchers have hypothesized that patterns of peer recognition change over time according to students' academic year, with gender biases in such recognition occurring in first-year, but not beyond first-year, science courses. Yet, this hypothesis has not been tested with data over a time-scale longer than one semester. In our study, we aimed to isolate the effect of time on peer recognition by examining a single cohort of students throughout their introductory physics sequence at a mostly-women, liberal arts institution. Results indicated that peer recognition is susceptible to change over time, but likely does not obey as universal a pattern as previously thought (i.e., patterns of peer recognition being explicitly tied to whether a course is aimed at first-year or beyond first-year students). Instead, patterns of recognition may change due to variables such as fluctuating student participation throughout a course sequence and, therefore, are likely course-specific. These findings suggest that interventions aimed at providing equitable opportunities for receiving peer recognition ought to be implemented consistently throughout a course sequence, rather than only early on as suggested by existing studies. Moreover, interventions to increase the number of students who verbally participate in whole-class discussions are likely more effective than those that broaden peer interaction networks. Such instructional efforts are important for ensuring that all students have the opportunity to feel recognized as a science person, develop their science identity, and persist in their science courses.

\vspace{0.4cm}

\section*{ACKNOWLEDGEMENTS}

This material is based upon work supported by the National Science Foundation Graduate Research Fellowship Program Grant No. DGE-2139899. We thank Natasha Holmes, Lee Simpfendoerfer, and Eric Brewe for meaningful feedback on this work.

\bibliography{brenau_perceptions.bib} 

\begin{thebibliography}{10}

\bibitem{gee2000chapter}
James~Paul Gee.
\newblock Chapter 3: Identity as an analytic lens for research in education.
\newblock {\em Review of Research in Education}, 25(1):99--125, 2000.

\bibitem{carlone2007understanding}
Heidi~B. Carlone and Angela Johnson.
\newblock Understanding the science experiences of successful women of color:
  Science identity as an analytic lens.
\newblock {\em Journal of Research in Science Teaching}, 44(8):1187--1218,
  2007.

\bibitem{hazari2010connecting}
Zahra Hazari, Gerhard Sonnert, Philip~M. Sadler, and Marie-Claire Shanahan.
\newblock Connecting high school physics experiences, outcome expectations,
  physics identity, and physics career choice: A gender study.
\newblock {\em Journal of Research in Science Teaching}, 47(8):978--1003, 2010.

\bibitem{hyater2018critical}
Simone Hyater-Adams, Claudia Fracchiolla, Noah Finkelstein, and Kathleen Hinko.
\newblock Critical look at physics identity: An operationalized framework for
  examining race and physics identity.
\newblock {\em Physical Review Physics Education Research}, 14(1):010132, 2018.

\bibitem{lock2013physics}
Robynne~M. Lock, Zahra Hazari, and Geoff Potvin.
\newblock Physics career intentions: The effect of physics identity, math
  identity, and gender.
\newblock In {\em AIP Conference Proceedings}, volume 1513, pages 262--265.
  American Institute of Physics, 2013.

\bibitem{hazari2018towards}
Zahra Hazari and Cheryl Cass.
\newblock Towards meaningful physics recognition: What does this recognition
  actually look like?
\newblock {\em The Physics Teacher}, 56(7):442--446, 2018.

\bibitem{hazari2017importance}
Zahra Hazari, Eric Brewe, Renee~Michelle Goertzen, and Theodore Hodapp.
\newblock The importance of high school physics teachers for female students’
  physics identity and persistence.
\newblock {\em The Physics Teacher}, 55(2):96--99, 2017.

\bibitem{kalender2019gendered}
Z.~Yasemin Kalender, Emily Marshman, Christian~D. Schunn, Timothy~J.
  Nokes-Malach, and Chandralekha Singh.
\newblock Gendered patterns in the construction of physics identity from
  motivational factors.
\newblock {\em Physical Review Physics Education Research}, 15(2):020119, 2019.

\bibitem{boe2023cleverness}
Maria~Vetleseter Bøe.
\newblock Staying recognised as clever: high-achieving physics students'
  identity performances.
\newblock {\em Physics Education}, 58(3):035012, 2023.

\bibitem{kalender2019female}
Z.~Yasemin Kalender, Emily Marshman, Christian~D. Schunn, Timothy~J.
  Nokes-Malach, and Chandralekha Singh.
\newblock Why female science, technology, engineering, and mathematics majors
  do not identify with physics: They do not think others see them that way.
\newblock {\em Physical Review Physics Education Research}, 15(2):020148, 2019.

\bibitem{grunspan2016}
Daniel~Z. Grunspan, Sarah~L. Eddy, Sara~E. Brownell, Benjamin~L. Wiggins,
  Alison~J. Crowe, and Steven~M. Goodreau.
\newblock Males under-estimate academic performance of their female peers in
  undergraduate biology classrooms.
\newblock {\em PloS One}, 11(2):e0148405, 2016.

\bibitem{salehi2019}
Shima Salehi, N.G. Holmes, and Carl Wieman.
\newblock Exploring bias in mechanical engineering students' perceptions of
  classmates.
\newblock {\em PloS One}, 14(3):e0212477, 2019.

\bibitem{bloodhart2020}
Brittany Bloodhart, Meena~M. Balgopal, Anne Marie~A. Casper, Laura~B
  Sample~McMeeking, and Emily~V. Fischer.
\newblock Outperforming yet undervalued: Undergraduate women in stem.
\newblock {\em PloS One}, 15(6):e0234685, 2020.

\bibitem{sundstrom2022perceptions}
Meagan Sundstrom, Ashley~B. Heim, Barum Park, and N.~G. Holmes.
\newblock Introductory physics students' recognition of strong peers: Gender
  and racial or ethnic bias differ by course level and context.
\newblock {\em Phys. Rev. Phys. Educ. Res.}, 18:020148, Dec 2022.

\bibitem{sundstromWhoWhat}
Meagan Sundstrom, L.~N. Simpfendoerfer, Annie Tan, Ashley~B. Heim, and N.~G.
  Holmes.
\newblock Who and what gets recognized in peer recognition.
\newblock arXiv preprint arXiv:2305.02415, 2023.

\bibitem{cid2020demographics}
Stephen Kanim and Ximena~C. Cid.
\newblock Demographics of physics education research.
\newblock {\em Phys. Rev. Phys. Educ. Res.}, 16:020106, Jul 2020.

\bibitem{lin2014peer}
Shih-Yin Lin, Scott~S Douglas, John~M Aiken, Chien-Lin Liu, Edwin~F Greco,
  Brian~D Thoms, Marcos~D Caballero, and Michael~F Schatz.
\newblock Peer evaluation of video lab reports in an introductory physics
  {MOOC}.
\newblock {\em arXiv preprint arXiv:1407.4714}, 2014.

\bibitem{zwolak2018educational}
Justyna~P. Zwolak, Michael Zwolak, and Eric Brewe.
\newblock Educational commitment and social networking: The power of informal
  networks.
\newblock {\em Physical Review Physics Education Research}, 14(1):010131, 2018.

\bibitem{traxler2020network}
Adrienne~L. Traxler, Tyme Suda, Eric Brewe, and Kelley Commeford.
\newblock Network positions in active learning environments in physics.
\newblock {\em Physical Review Physics Education Research}, 16(2):020129, 2020.

\bibitem{dou2019practitioner}
Remy Dou and Justyna~P. Zwolak.
\newblock Practitioner’s guide to social network analysis: Examining physics
  anxiety in an active-learning setting.
\newblock {\em Physical Review Physics Education Research}, 15(2):020105, 2019.

\bibitem{commeford2021characterizing}
Kelley Commeford, Eric Brewe, and Adrienne Traxler.
\newblock Characterizing active learning environments in physics using network
  analysis and classroom observations.
\newblock {\em Physical Review Physics Education Research}, 17(2):020136, 2021.

\bibitem{sundstrom2022interactions}
Meagan Sundstrom, Andy Schang, Ashley~B. Heim, and N.~G. Holmes.
\newblock Understanding interaction network formation across instructional
  contexts in remote physics courses.
\newblock {\em Phys. Rev. Phys. Educ. Res.}, 18:020141, Dec 2022.

\bibitem{smith2013structural}
Jeffrey~A. Smith and James Moody.
\newblock Structural effects of network sampling coverage i: Nodes missing at
  random.
\newblock {\em Social networks}, 35(4):652--668, 2013.

\bibitem{grunspan2014}
Daniel~Z. Grunspan, Benjamin~L. Wiggins, and Steven~M. Goodreau.
\newblock Understanding classrooms through social network analysis: A primer
  for social network analysis in education research.
\newblock {\em CBE—Life Sciences Education}, 13(2):167--178, 2014.

\bibitem{brewe2018guide}
Eric Brewe.
\newblock The roles of engagement: Network analysis in physics education
  research.
\newblock In {\em Getting Started in PER}, volume~2. American Association of
  Physics Teachers, College Park, MD, July 2018.

\bibitem{butts2006exact}
Carter~T. Butts.
\newblock Exact bounds for degree centralization.
\newblock {\em Social Networks}, 28(4):283--296, 2006.

\bibitem{Tesch1990}
R.~Tesch.
\newblock Qualitative research—analysis types and software protocols.
\newblock {\em Hampshire, UK: The Falmer Press}, 1990.

\bibitem{buchert2008first}
Stephanie Buchert, Eric~L. Laws, Jennifer~M. Apperson, and Norman~J. Bregman.
\newblock First impressions and professor reputation: Influence on student
  evaluations of instruction.
\newblock {\em Social Psychology of Education}, 11:397--408, 2008.

\bibitem{laws2010student}
Eric~L. Laws, Jennifer~M. Apperson, Stephanie Buchert, and Norman~J. Bregman.
\newblock Student evaluations of instruction: When are enduring first
  impressions formed?
\newblock {\em North American Journal of Psychology}, 12(1), 2010.

\bibitem{eddy2014}
Sarah~L. Eddy, Sara~E. Brownell, and Mary~Pat Wenderoth.
\newblock Gender gaps in achievement and participation in multiple introductory
  biology classrooms.
\newblock {\em CBE—Life Sciences Education}, 13(3):478--492, 2014.

\bibitem{eddy2015caution}
Sarah~L. Eddy, Sara~E. Brownell, Phonraphee Thummaphan, Ming-Chih Lan, and
  Mary~Pat Wenderoth.
\newblock Caution, student experience may vary: Social identities impact a
  student’s experience in peer discussions.
\newblock {\em CBE—Life Sciences Education}, 14(4):ar45, 2015.

\bibitem{aguillon2020gender}
Stepfanie~M. Aguillon, Gregor-Fausto Siegmund, Renee~H. Petipas, Abby~Grace
  Drake, Sehoya Cotner, and Cissy~J. Ballen.
\newblock Gender differences in student participation in an active-learning
  classroom.
\newblock {\em CBE—Life Sciences Education}, 19(2):ar12, 2020.

\bibitem{nichols2022participation}
Sierra~C. Nichols, Yongyong~Y. Xia, Mikaylie Parco, and Elizabeth~G. Bailey.
\newblock Participation and performance by gender in synchronous online
  lectures: three unique case studies during emergency remote teaching.
\newblock {\em Journal of Microbiology \& Biology Education}, 23(1):e00281--21,
  2022.

\bibitem{tanner2013structure}
Kimberly~D. Tanner.
\newblock Structure matters: twenty-one teaching strategies to promote student
  engagement and cultivate classroom equity.
\newblock {\em CBE—Life Sciences Education}, 12(3):322--331, 2013.

\end{thebibliography}

\section{Appendix A: Survey format}

\begin{figure*}[t]
    \centering
    \includegraphics[width=6.2in,trim={0 0 3cm 0}]{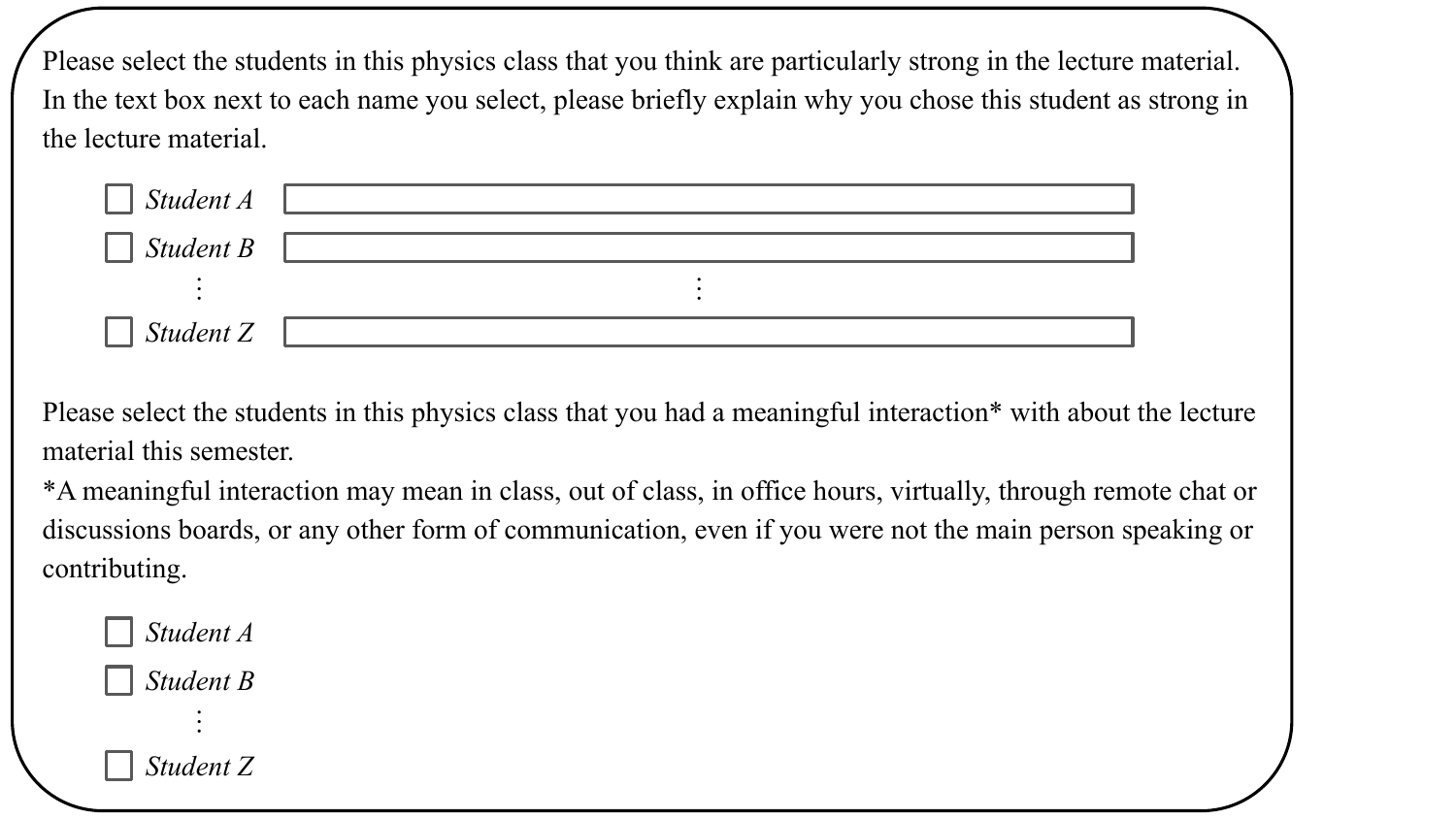}
    \caption{Setup of the network survey administered to students online via Qualtrics. Small boxes on the left of student names are check boxes. Large boxes to the right of student names are open text boxes.
    }
    \label{fig:survey}
\end{figure*}

Figure~\ref{fig:survey} shows the setup of the network survey administered to students. 

\end{document}